\documentclass[a4paper,11pt]{article}
\pdfoutput=1
\usepackage{jcappub}
\usepackage{mathrsfs}
\usepackage{bm}
\usepackage[marginal]{footmisc}
\usepackage{aas_macros}
\usepackage{graphicx}
\usepackage{dcolumn}
\usepackage{bm}
\usepackage{float}
\usepackage{wrapfig}
\usepackage{subfigure}
\usepackage{booktabs}
\usepackage{url}
\usepackage{appendix}

\newcommand{\be}{\begin{equation}}
	\newcommand{\ee}{\end{equation}}
\newcommand{\ba}{\begin{eqnarray}}
	\newcommand{\ea}{\end{eqnarray}}

\begin{document}
	\title{Constraints on ultra-slow-roll inflation with the NANOGrav 15-Year Dataset}

\author[a,b]{Bo Mu,}
\author[a,b]{Jing Liu,}
\author[a,b]{Gong Cheng,}
\author[c,d,e]{Zong-Kuan Guo,}

\affiliation[a]{
	International Centre for Theoretical Physics Asia-Pacific, University of Chinese Academy of Sciences, 100190 Beijing, China
}
\affiliation[b]{
	Taiji Laboratory for Gravitational Wave Universe, University of Chinese Academy of Sciences, 100049 Beijing, China
}
\affiliation[c]{
    CAS Key Laboratory of Theoretical Physics, Institute of Theoretical Physics, Chinese Academy of Sciences, P.O. Box 2735, Beijing 100190, China
}
\affiliation[d]{
    School of Physical Sciences, University of Chinese Academy of Sciences, No.19A Yuquan Road, Beijing 100049, China}
\affiliation[e]{
    School of Fundamental Physics and Mathematical Sciences, Hangzhou Institute for Advanced Study, University of Chinese Academy of Sciences, Hangzhou 310024, China
}

\emailAdd{mubo22@mails.ucas.ac.cn}
\emailAdd{liujing@ucas.ac.cn}
\emailAdd{chenggong@ucas.ac.cn}
\emailAdd{guozk@itp.ac.cn}

\abstract{
    Ultra-slow-roll~(USR) inflation  predicts an exponential amplification of scalar perturbations at small scales,
    which leads to a stochastic gravitational wave background~(SGWB) through the coupling of the scalar and tensor modes at the second-order expansion of the Einstein equation.
    In this work, we search for such a scalar-induced SGWB from the NANOGrav 15-year (NG15) dataset,
    and find that the SGWB from USR inflation could explain the observed data.
    The Bayes factors are $54\pm 5$ for the USR inflation model alone and $68\pm 6$ for the combined USR inflation plus supermassive black hole binaries (SMBHB) models.
    We place constraints on the amplitude of the scalar power spectrum to $P_{\mathrm{Rp}} > 10^{-1.95}$ at $95\%$ confidence level (C.L.) at the scale of $k\sim 20\, \mathrm{pc}^{-1}$.
    We find that $\log_{10} P_{\mathrm{Rp}}$ degenerates with the peak scale $\log_{10} k_{\mathrm{p}}$.
    We also obtain the parameter space allowed by the data in the USR inflationary scenario, where the $e$-folding numbers of the duration of the USR phase has a lower limit $\Delta N > 2.80$ ($95\%$ C.L.) when the USR phase ends at $N\approx 20$.
    With astrophysically motivated priors, the NG15 dataset fits both the USR inflation model and SMBHB model equally well.
    
}
	
	\maketitle
	\section{Introduction}
The direct detection of gravitational waves~(GWs), which are the ripples of spacetime predicted by general relativity, opens a new era of the exploration of the early Universe~\cite{abbott2016observation,Cai:2017cbj,Bian:2021ini}. The monumental discovery was the detection of the merger of two black holes by LIGO which provides a new way to observe the Universe, complementing traditional electromagnetic observations.
SGWBs are formed from the superpositions of many unresolved GW sources, of both astrophysical and cosmological origins, which can be detected by searching for correlated signals between multiple detectors. Searching for such SGWBs has been the main scientific goal of multiband GW observers including LIGO-Virgo-KAGRA, LISA, Taiji, TianQin, and pulsar timing array~(PTA) experiments.

Recently, the PTA experiments, including CPTA~\cite{Xu:2023wog}, NANOGrav~\cite{NANOGrav:2023hvm}, PPTA~\cite{Reardon:2023gzh} and EPTA~\cite{Antoniadis:2023ott}, collectively announced the first positive evidence of SGWBs at the nanohertz frequency band ($\sim 1-100\,\mathrm{nHz}$).
Based on the evolution of the Universe, it is predicted that a cosmic population of supermassive black hole binaries generates an SGWB at the nanohertz frequency band, which is treated as the fiducial model in the analysis of NANOGrav-15 dataset~\cite{2306.16213}. However, the detected GW power spectrum deviates from that given by the expected values of each parameter~\cite{NANOGrav:2023hfp}. Actually there are also other potential GW sources at this frequency band, such as cosmological phase transitions~\cite{Han:2023olf,Jiang:2023qbm,Fujikura:2023lkn}, cosmic topological defects~\cite{Ellis:2023tsl,Wang:2023len,Lazarides:2023ksx,Kitajima:2023cek,Gouttenoire:2023ftk}, and primordial scalar perturbations~\cite{Chen:2019xse,Madge:2020jjt,Inomata:2023zup,Liu:2023ymk,You:2023rmn,Wang:2023ost,Zhao:2023joc,Basilakos:2023xof,Basilakos:2023jvp,Jin:2023wri,Liu:2023pau,Liu:2023hpw,Firouzjahi:2023lzg,Franciolini:2023pbf,Ellis:2023oxs,Choudhury:2023fwk,Choudhury:2023hfm}, which generate SGWBs in the early Universe. It is found that some of these cosmological sources can explain the NANOGrav dataset~\cite{NANOGrav:2023hvm}.
In their work, three parameterization templates for the power spectrum of scalar perturbations, which source the energy spectrum of scalar-induced gravitational waves (SIGWs), have been considered.

In this paper, we fit the NG15 dataset with the SGWB from ultra-slow-roll (USR) inflation (marked as SIGW-USR in this paper), which is a physically well-motivated SIGW model.
In the USR inflationary scenario, where the slow-roll approximation is not adhered to, the superhorizon evolution of scalar perturbations can lead to exponential amplification as calculated in various studies~\cite{Liu:2020oqe,Ozsoy:2019lyy,Byrnes:2018txb}. The USR regime arises from the non-attractor evolution of the inflaton field, which can be manifested through modified gravity theories~\cite{Pi:2022zxs,Lin:2020goi,Yi:2022anu,Kawai:2021edk,Kawai:2021bye}, string theory approaches~\cite{Cicoli:2018asa,Cicoli:2022sih,Ozsoy:2018flq}, and supergravity models~\cite{Dalianis:2018frf,Gao:2018pvq,Wu:2021zta,Kawai:2022emp}. Additionally, minor fluctuations in the inflationary potential -- encompassing features like bumps~\cite{Mishra:2019pzq,Ozsoy:2020kat}, dips~\cite{Gu:2022pbo,Mishra:2019pzq}, inflection points~\cite{Choudhury:2013woa,Germani:2017bcs,Bhaumik:2019tvl,Bhaumik:2019tvl}, and steps~\cite{Kefala:2020xsx,Inomata:2021uqj,Cai:2021zsp,Inomata:2021tpx} -- might also be considered as possible triggers for the USR regime. A myriad of references~\cite{Carr:2016drx,Bird:2016dcv,Di:2017ndc,Garcia-Bellido:2017mdw,Hertzberg:2017dkh,Passaglia:2018ixg,Cai:2018dig,Fu:2020lob,Cai:2022erk,Figueroa:2021zah,Figueroa:2020jkf,Pi:2021dft,Wang:2021kbh,Xu:2019bdp,Cheong:2019vzl,Braglia:2020eai} suggest that enhanced curvature perturbations might pave the way for the emergence of primordial black holes housing dark matter, rendering USR inflation a subject of keen interest. Based on cosmic microwave background~(CMB) observations, the amplitude of the primordial scalar power spectrum is firmly constrained to be around $~ 2.10\times 10^{-9}$\cite{Planck:2018vyg}. Nonetheless, at smaller scales, these constraints might be more flexible~\cite{Emami:2017fiy,Gow:2020bzo,Cyr:2023pgw}. The USR domain can also yield significant $e$-folding numbers, crucial parameters for addressing horizon, flatness, and monopole issues~\cite{Pattison:2018bct}. The core objective of our research is to discern the constraints that the NG15 dataset might impose on the inflationary potential of the USR regime as well as on the power spectrum of scalar perturbations.
For convenience, $c=8\pi G=1$ is set throughout this paper.

	\section{PTA Data}
	We use the NG15 dataset, which comprises the pulse time of arrival (ToAs) for 68 millisecond pulsars. With a timing baseline of 3 years, 67 of these pulsars remain viable for processing. NANOGrav fits these ToAs to a timing model that encapsulates the sky location, proper motion, parallax, pulsar spin period, and spin period derivative for each pulsar~\cite{2306.16213}.

    We use the Python package \texttt{PTArcade}~\cite{lamb2023need} to perform the Bayesian analysis, which integrates new physics into the PTA data analysis package \texttt{ENTERPRISE}~\cite{enterprise}. New physics is characterized by the GW spectrum, denoted as \(\Omega_{\mathrm{GW}}\). Except for \(\Omega_{\mathrm{GW}}\), we set other parameters by default, such as the ephemeris model included in the timing model.

    Then, we will briefly introduce the likelihood function used in this work. The pulsar's timing residuals $\bm{\delta t}$ can be modeled as
     \begin{equation}
         \label{defination of delta t}
         \bm{\delta t}= \bm{n}+\bm{M} \bm{\epsilon}+\bm{F} \bm{a}
         \,,
     \end{equation}
     where \(\bm{n}\) describes the white noise, \(\bm{M} \bm{\epsilon}\) represents the errors associated with the best-fitting timing-ephemeris parameters~\cite{Vallisneri_2020}. The term \(\bm{F}\bm{a}\) characterizes the red noise, which is presumed to be a combination of pulsar-intrinsic red noise and SGWB signals. The matrix \(\bm{F}\) is the Fourier basis matrix, constructed from sine-cosine pairs based on the ToAs with frequencies defined as \(f_i = i/T_{\mathrm{obs}}\), where \(T_{\mathrm{obs}} = 16.03\text{yr}\) is the timing baseline for the entire NG15 dataset. For the Fourier amplitudes, \(\bm{a}\) follows a zero-mean normal distribution with a covariance matrix  $ \langle \bm{a}\bm{a}^T\rangle=\bm{\phi}$, which is given by
     \begin{equation}
         \label{covatriance matrix of a}
         \left[\phi\right]_{\left(ak\right)\left(bj\right)} =\delta_{ij}\left(\delta_{ab}\varphi_{a,i}+\Gamma_{ab}\Phi_i  \right)
         \,.
     \end{equation}
    The first term characterizes the pulsar-intrinsic red noise, which is defined as a power-law function in this work, with the coefficients \(\varphi_{a,i}\) modeled as
     \begin{equation}
         \label{varphi}
         \varphi_{a,i}\left(f\right)=\frac{A^2_a}{12\pi^2}\frac{1}{T_{\mathrm{obs}}}\left(\frac{f}{1 \mathrm{yr}^{-1}}\right)^{-\gamma_a}\mathrm{yr}^{3}
         \,,
     \end{equation}
    and the priors of \(A_a,\gamma_a\) are shown in Table~I. In the second term, \(\Gamma_{ab}\) is defined based on the Hellings \& Downs ("HD") correlation~\cite{1983ApJ...265L..39H}. \(\Phi_i\) is related to the SGWB spectrum as
     \begin{equation}
         \label{Omega GW to Phi}
         \Omega_{\mathrm{GW}}\left(f\right)\equiv \frac{1}{\rho_c}\frac{d\rho _{\mathrm{GW}}\left(f\right)}{d\mathrm{ln}\left(f\right)}=\frac{8\pi^4 f^5}{H_0^2}\frac{\Phi\left(f\right)}{\Delta f}
         \,.
     \end{equation}
     Here $H_0= h\times 100~\mathrm{km~s}^{-1}\mathrm{Mpc}^{-1}$ is the Hubble constant, $\Delta f=1/T_{\mathrm{obs}}$, and $\Phi\left(f\right)$ is defined as $\Phi_i=\Phi\left(i/T_{\mathrm{obs}}\right)$.

     Upon marginalizing over \(\bm{a},\bm{\epsilon}\), we obtain a likelihood that depends only on the parameters that affect \(\langle \bm{a}\bm{a}^T\rangle\). The likelihood is expressed as
     \begin{equation}
         \label{likelihood function}
         p\left(\bm{\delta t}|\bm{\phi}\right)=\frac{\mathrm{exp}\left(-\frac{1}{2}\bm{\delta t}^T \bm{C}^{-1}\bm{\delta t}\right)}{\sqrt{\mathrm{det}\left(2\pi \bm{C}\right)}}
         \,,
     \end{equation}
     where \(\bm{C} = \bm{N} + \bm{TBT}^T\), \(\bm{N}\) is the covariance matrix of \(\bm{n}\), \(\bm{T} = [\bm{M},\bm{F}]\), and \(\bm{B} = \mathrm{diag}(\bm{\infty},\bm{\phi})\). Here, \(\bm{\infty}\) represents a diagonal matrix filled with infinities, implying that the priors for parameters in \(\bm{\epsilon}\) are assumed to be flat.

	\section{SGWB from USR inflation}
Under the slow-roll approximation, the Fourier modes of curvature perturbations, \(R_k\), remain constant at superhorizon scales. Refs.~\cite{Liu:2020oqe,Ozsoy:2019lyy,Byrnes:2018txb}  find that during the USR regime at superhorizon scales, the time derivative of \(R_k\), denoted as \(\dot{R}_k\), undergoes exponential amplification. Consequently, the primordial scalar power spectrum \(P_{R}\) derived from \(R_k\) reaches its peak at the beginning of USR inflation. Ref.~\cite{Liu:2020oqe} suggests that \(P_{\mathrm{R}}\) behaves as \(P_{\mathrm{R}} \propto k^4\) on the infrared side of the peak, and \(P_{\mathrm{R}} \propto k^\beta\) on the ultraviolet side, with \(\beta\) depending on the inflationary potential. Given that USR inflation ends at \(\phi=\phi_{\mathrm{e}}\), the Taylor expansion of the inflationary potential around \(\phi_{\mathrm{e}}\) is
\begin{equation}
	    \label{eq:potential energy }
	    V\left(\phi\right)=b_0+b_1\left(\phi-\phi_{\mathrm{e}}\right)+b_2\left(\phi-\phi_{\mathrm{e}}\right)^2+\cdots
     \,.
\end{equation}
In a Friedmann-Robertson-Walker universe, the Friedmann equation and the EOM of $\phi$ are written as
\begin{equation}
\begin{aligned}
	\label{eq:Friedmann equation and the EOM of phi}
	H^2=\frac{1}{3}\left(\dot{\phi}^2+V\left(\phi\right)\right),\\
    \ddot{\phi}+3H\dot{\phi}+\frac{dV}{d\phi}=0,
\end{aligned}
\end{equation}
and with $|\dot{\phi}|$ reaches its minimum at $t_e$, we can obtain
\begin{equation}
\begin{aligned}
	\label{eq:middle terms for b_1=0}
	\ddot{\phi}(t_e)=&0,\\
    3H\dot{\phi}(t_e)=&-\frac{dV}{d\phi}(t_e)=-b_1,
\end{aligned}
\end{equation}
With more detailed assumptions and further calculation~\cite{Liu:2020oqe}, one can obtain
 \begin{equation}
	    \label{eq:param beta}
	    \beta=3-\sqrt{9-24b_2/b_0}
     \,,
\end{equation}
which can be treated as a constant as long as $\phi$ is in the vicinity of $\phi_{\mathrm{e}}$. During the USR regime $\dot{\phi}$ exponentially decreases, after which  $\phi$ stays close to $\phi_{e}$ for a long period during inflation so that the expansion~\eqref{eq:potential energy } remains valid. Due to the smallness of $\dot{\phi}(t_{e})$, the parameter $b_{1}$ should also be very small. Note that the spectral index of the amplified $P_{R}(k)$ is independent of the linear term in Eq.~\eqref{eq:potential energy }. The term $\dot{\phi}$ exponentially decreases in the short USR regime, reaching its minimum at $t_{e}$. Subsequently, $\dot{\phi}$ needs to increase significantly to ensure a successful end of inflation. Apart from the rather short period around $t_{e}$, the term $\dot{\phi}$ is orders of magnitude larger than its minimum value, leading to the dominance of the dynamics of $\phi$ and the profile of $P_{R}(k)$ by the quadratic term, $b_{2}(\phi-\phi_{e})^{2}$, in $V(\phi)$.

Instead of numerically calculating the exact form of \(P_{\mathrm{R}}(k)\), we parameterize the power spectrum of curvature perturbations in USR inflation as
	\begin{equation}
	    \label{eq:defination of PRk}
	    P_{\mathrm{R}}\left(k\right)=P_{R\mathrm{p}}\cfrac{\left(\alpha+\beta\right)^\gamma}{\left[\beta \left(k/k_{\mathrm{p}}\right)^{-\alpha/\gamma}+\alpha \left(k/k_{\mathrm{p}}\right)^{\beta/\gamma}\right]^\gamma}\,,
	\end{equation}
	where \(P_{\mathrm{R}}(k)\) reaches its peak value \(P_{R\mathrm{p}}\) at \(k=k_{\mathrm{p}}\), and we set \(\alpha = 4\) following the relation \(P_{\mathrm{R}} \propto k^4\). The parameter \(\gamma\) characterizes the smoothness of \(P_{\mathrm{R}}(k)\) around the peak. Since \(\gamma\) does not significantly affect the results, we adopt a representative value of \(\gamma = 2.6\) in this work. The justification for this selection is detailed in the appendix. Then, the remaining free parameters are \(P_{R\mathrm{p}}\), \(k_{\mathrm{p}}\), and \(\beta\).

	Given the form of $P_{\mathrm{R}}\left(k\right)$, we are able to numerically calculate the GW energy spectrum $\Bar{\Omega}_{\mathrm{GW}}\left(k\right)$ in the radiation-dominated era, following the integration method in Refs.~\cite{Kohri:2018awv,Espinosa:2018eve}
	\begin{equation}
    \begin{aligned}
    \Bar{\Omega}_{\mathrm{GW}}\left(k\right)= \int_0^{\infty}dv \int_{|1-v|}^{1+v}du \left(\frac{4v^2-\left(1+v^2-u^2\right)^2}{4uv}\right)^2 IRD_{sq}\left(u,v\right)P_{\mathrm{R}}\left(ku\right)P_{\mathrm{R}}\left(kv\right)
    \,,
    \end{aligned}	
	\end{equation}
 in which $IRD_{sq}\left(u,v\right)$ is a function of $u,v$,
 \begin{equation}
     \label{IRD equation}
     \begin{aligned}
     IRD_{sq}(u,v)= &
     \frac{1}{2}\left(\frac{3(u^2+v^2-3)^2}{4u^3v^3}\right)^2\times \Bigg[\left(-4uv  +\left(u^2+v^2-3\right)\mathrm{ln}\left|\frac{3-(u+v)^2}{3-(u-v)^2}\right|\right)^2\\
     &+\pi^2\left(u^2+v^2-3\right)^2\Theta\left(u+v-\sqrt{3}\right)\Bigg]
     \,,
     \end{aligned}
 \end{equation}
 where $\Theta(x)$ is the heaviside step function.

 Furthermore, using the relationship \(f \approx 0.03\,\text{Hz} \frac{k}{2 \times 10^7 \, \text{pc}^{-1}}\), we could express $\Bar{\Omega}_{\mathrm{GW}}(k)$ in terms of the corresponding frequency \(f\). Repeatedly calculating this integration for every step would be computationally expensive. As the integral is simply proportional to $P_{\mathrm{Rp}}^{2}$, we extract $P_{\mathrm{Rp}}^{2}$ from the integral. So now the integral only depends on the other two parameters \(\beta\) and \(f_{\rm p}\). Then we compute the integration in the \(\beta\)-\(f_{\rm p}\) grid. So for any given values of \(\beta\) and \(f_{\rm p}\), we can infer the integral by using 2-D interpolation.

The energy spectrum of SGWB at present $\Omega_{\mathrm{GW}}(f)$ is related to $\Bar{\Omega}_{\mathrm{GW}}\left(f\right)$ as

\begin{equation}
	    \label{eq:param}
	    \Omega_{\mathrm{GW}}\left(f\right)=\Omega_r\left(\frac{g_*\left(f\right)}{g_*^0}\right)\left(\frac{g_{*,s}^0}{g_{*,s}\left(f\right)}\right)^{4/3}\Bar{\Omega}_{\mathrm{GW}}\left(f\right)
     \,,
\end{equation}
where $\Omega_{r}$ is the present energy density fraction of radiation, $g_{*}$ and $g_{*,s}$ respectively represent the effective relativistic degree of freedom that contribute
to the radiation energy and entropy density, the superscript $0$ denotes the present time.
To determine \(\Omega_{\mathrm{GW}}(f)\), we set the values \(\Omega_r/g_*^0 \approx 2.72 \times 10^{-5}\), \(g_{*,s}^0 \approx 3.93\), and the functions \(g_*(f)\) and \(g_{*,s}(f)\) given in Ref.~\cite{Saikawa_2020}.

 The priors adopted in this work are listed in Table I. By definition, the peak value \(P_{R\mathrm{p}}\) should not exceed \(\mathcal{O}\left(1\right)\) so that $R_{k}$ remains at the level of perturbations. Thus we set $\mathrm{log}_{10}P_{R\mathrm{p}}\leq 0$ to explore what constraints the NG15 dataset would impose on the USR model. Curvature perturbations with large amplitude are also responsible to the production of primordial black holes~(PBHs). To avoid the overproduction of PBHs, we also apply a conservative upper bound $\mathrm{log}_{10}P_{R\mathrm{p}}\leq -2$ for Gaussian perturbations.~\cite{Franciolini:2023pbf} There is no prior information for $f_{\mathrm{p}}$ since the USR regime can occur at any period during inflation. Therefore, we set \(f_{\mathrm{p}}\) to be roughly within the sensitivity band of NANOGrav. The energy spectrum in our model is a convex upward curve. If \(f_{\mathrm{p}}\) is below the band of NANOGrav, the spectrum can not fit the data. But if \(f_{\mathrm{p}}\) is beyond the band of NANOGrav, our model can still fit the data well. So we set the upper bound of the prior for \(f_{\mathrm{p}}\) two orders of magnitude higher than the band of NANOGrav. For the parameter \(\beta\), theoretically it can span a large range, so we set the prior of \(\beta\) between 0 and 5 to study its behavior. As the GWB from the mergers of SMBHBs is a promising signal in the PTA band, we explore the possibility of using a combined signal model that incorporates both SIGW-USR model and SMBHB model (SIGW-USR+SMBHB) to explain the NANOGrav dataset. The spectrum of SMBHB takes the form similar to Eq.(\ref{Omega GW to Phi}), with $\Phi(f)$ similar to Eq.(\ref{varphi}). The priors of SMBHB are set by the Python package \texttt{PTArcade} by default (see the appendix of Ref.~\cite{NANOGrav:2023hvm} for details). We list the parameter values for the 2-D Gaussian prior of the SMBHB model below.

\begin{eqnarray}
\label{2-D Gaussian parameters}
    \boldsymbol{\mu}_{\mathrm{BHB}}=
    \begin{pmatrix}
        -15.6\\
        4.7
    \end{pmatrix}
    ,
    \boldsymbol{\sigma}_{\mathrm{BHB}}=
    \begin{pmatrix}
        2.8 & -0.026\\
        -0.026 & 1.2
    \end{pmatrix}
\end{eqnarray}

	\begin{table}
		\begin{center}
			\caption{\label{table:prior} Priors on the Model Parameters. The first five parameters correspond to our USR model, while the 2-D Gaussian prior is applied to the SMBHB model. The exact values for $\boldsymbol{\mu}_{\mathrm{BHB}}$ and $\boldsymbol{\sigma}_{\mathrm{BHB}}$ are introduced in Eq.(\ref{2-D Gaussian parameters}). For $P_{R\mathrm{p}}$, we impose an upper limit of $\mathrm{log}_{10}P_{R\mathrm{p}}\leq 0$, disregarding astrophysical constraints, to explore what constraints the NG-15 dataset would impose on the USR model. Additionally, a more stringent upper limit of $\mathrm{log}{10}P_{R\mathrm{p}}\leq -2$ is set based on PBH abundance, allowing for comparison with the SMBHB model under astrophysically motivated priors.}



			\begin{tabular}{cc}
				\hline \hline
				Parameters& Priors \\
				\hline
				$P_{R\mathrm{p}}$ & Log-Uniform$\left(-3,-2/0\right)$ \\
				$f_{\mathrm{p}}$ $\left[\mathrm{Hz}\right]$  & Log-Uniform$\left(-10,-5\,\right)$ \\
				$\alpha$ & 4\\
				$\beta$ & Uniform$\left(0,5\right) $\\
				$\gamma$ & 2.6\\
                ($\mathrm{log}_{10}A_{\mathrm{BHB}}$,$\gamma_{\mathrm{BHB}}$) & Normal$\left(\boldsymbol{\mu}_{\mathrm{BHB}},\boldsymbol{\sigma}_{\mathrm{BHB}}\right)$\\

				\hline \hline
			\end{tabular}
		\end{center}
	\end{table}

	\section{Results}
	\begin{figure}[h]
 \centering
		\includegraphics[width=0.5\textwidth]{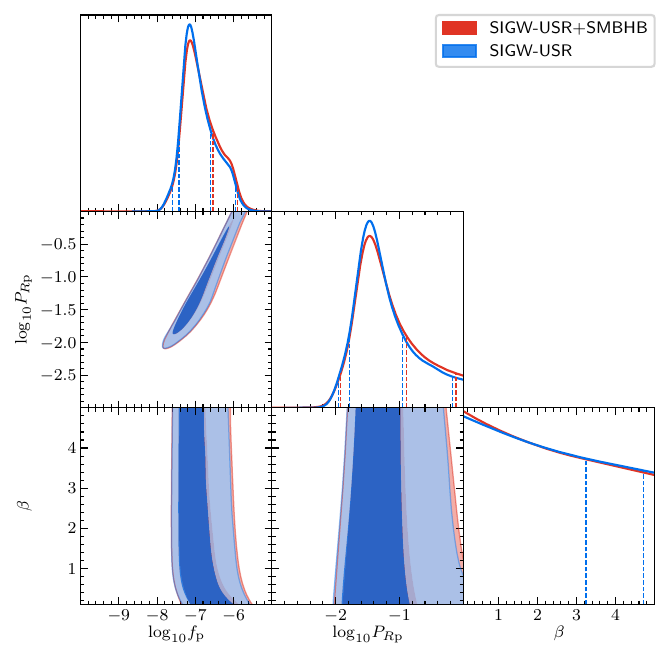}
		\caption{Posterior distribution of three input parameters. The red and blue contours denote the $68\%$ and $95\%$ confidence regions of SIGW-USR and SIGW-USR+SMBHB, respectively.
  } 
		\label{fig:posterior of usr modeled GW}
	\end{figure}
 We use the SGWB energy spectrum from SIGW-USR and the combined model SIGW-USR+SMBHB to fit the NG15 dataset and obtain the constraints on the model parameters. Fig.~\ref{fig:posterior of usr modeled GW} shows the posterior distributions for the three model parameters. We also calculate the Bayes factors against the baseline SMBHB model, yielding $\mathcal{B}=54\pm5$ for SIGW-USR and $\mathcal{B}=68\pm6$ for SIGW-USR+SMBHB. To compare with the Bayes factors obtained by other new physics models, we use the same prior for $P_{R\mathrm{p}}$ when calculating these two Bayes factors as in the SIGW models mentioned in Ref.~\cite{NANOGrav:2023hvm}, which follows a log-uniform distribution $\left(-3,1\right)$. The Bayes factors are close to those for the SIGW-Gauss model, the best-fitting model for the NG15 dataset reported in Ref.~\cite{NANOGrav:2023hvm}.  This suggests the SIGW-USR model also provides a good fit for the NG15 dataset. 
 
 As the PTA data points are roughly monotonically increasing, in the posterior plot, the peak of $\Omega_{\rm GW}$ will degenerate with \(f_{\mathrm{p}}\). And the peak of $\log_{10} \Omega_{\rm GW}$ is proportional to \( \log_{10} P_{\mathrm{Rp}}\), so we find that in the \( \log_{10} P_{\mathrm{Rp}} -\log_{10} f_{\mathrm{p}} \) plot, the two parameters degenerate with each other. The degeneracy could also be found in the power-law energy spectrum model. The lower limit of \( \log_{10} P_{\mathrm{Rp}} \) is approximately\(-1.95\) at \(95\%\) C.L. for  SIGW-USR and SIGW-USR+SMBHB. In the one-dimensional marginalized distribution, the peak values of \( \log_{10} P_{\mathrm{Rp}} \) and \(\log_{10} f_{\mathrm{p}} \) are $-1.46$ and $-7.14$ for SIGW-USR, $-1.47$ and $-7.14$ for SIGW-USR+SMBHB, respectively. Besides, the parameter \(\beta\) is poorly constrained as expected.

  	\begin{figure}[h]
				\includegraphics[width=0.5\textwidth]{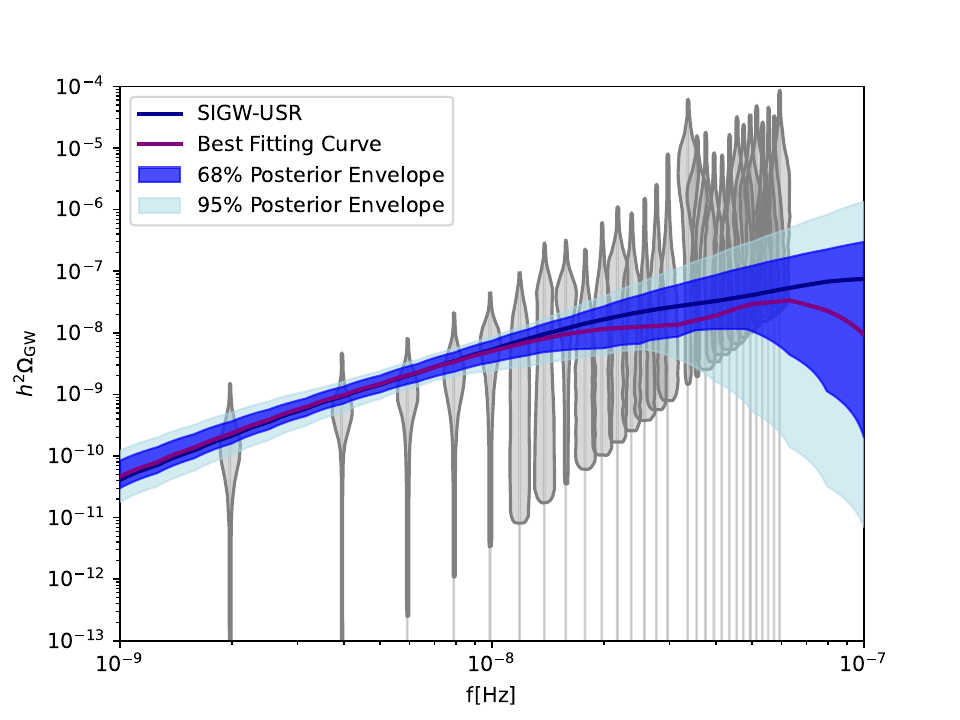}
        \includegraphics[width=0.5\textwidth]{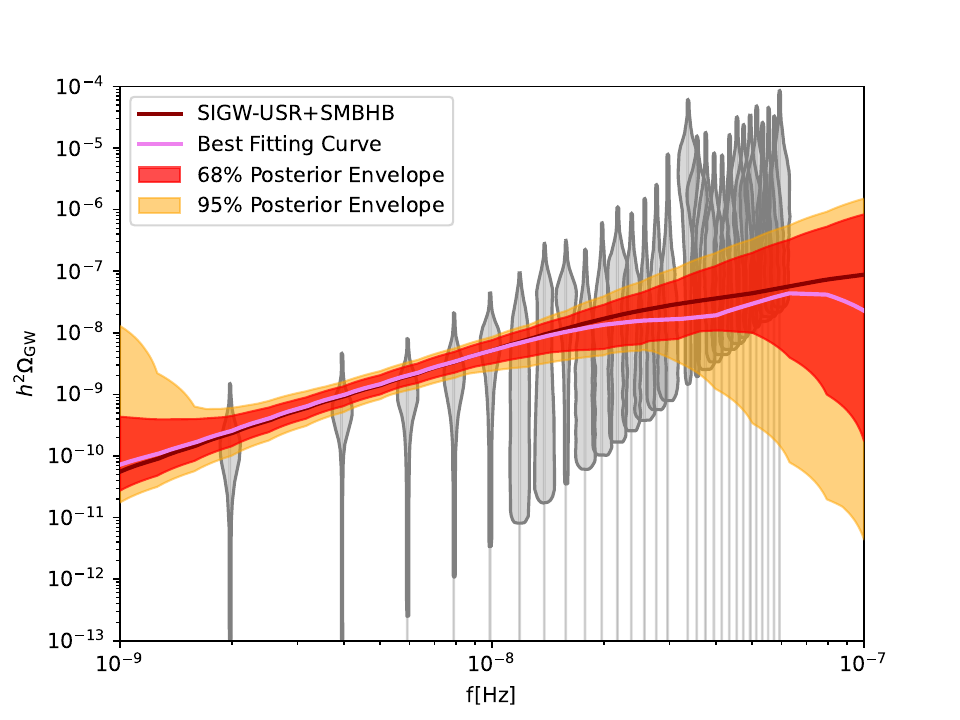}
		\caption{The $68\%$ and $95\%$ posterior envelopes, the best-fitting SGWB power spectra, and the median SGWB power spectra for both the SIGW-USR and SIGW-USR+SMBHB models are shown. The label of the median GW power spectra is used to indicate the corresponding model. The gray violin plots represent the periodogram for the free spectral process from the NANOGrav dataset~\cite{NANOGrav:2023hvm}. }
		\label{fig:USR modeled GW with NANOGrav data}
	\end{figure}

In Fig.~\ref{fig:USR modeled GW with NANOGrav data}, we present The $68\%$ and $95\%$ posterior envelopes, the best fitting SGWB power spectra and the median GW power spectra of both SIGW-USR and SIGW-USR+SMBHB models, and we plot the NANOGrav data points~\cite{NANOGrav:2023hvm} as well. The energy spectrum in our model is a convex upward curve. So to fit the data, \(f_{\mathrm{p}}\) should be at or above the NANOGrav band. The infrared spectrum of our model is governed by the spectrum index \( \alpha \) which is a constant in our model. The UV spectrum of our model is governed by the index \( \beta \). The constraining power of NANOGrav data mostly comes from the low-frequency and intermediate-frequency data points, so \( \beta \) is loosely constrained. 


 \begin{figure}[h]
        \centering
		\includegraphics[width=0.6\textwidth]{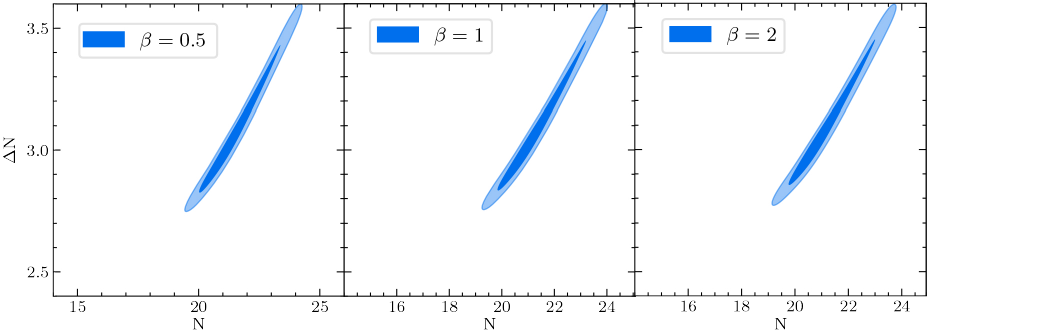}
		\caption{The confidence contours of $N-\Delta N$ with $\beta$ set as $0.5,1$ and $2$ from left to right.}
		\label{fig:N-DeltaN posterior}
	\end{figure}
In Fig.~\ref{fig:N-DeltaN posterior}, we present our constraints on \(N\) and \(\Delta N\), the intrinsic parameters of USR inflation, where \(N\) and \(\Delta N\) respectively denotes the \(e\)-folding number of the end and the duration of the USR regime~\footnote{The exact definition of $N$ is $N\equiv\ln[a(t_{e})/a(t_{i})]$, where $a$ is the scale factor, $t_{i}$ is the time when the Hubble scale at present leaves the horizon during inflation, and $t_{e}$ is the end of the USR regime.}. Through an analytical calculation to predict \(P_{\mathrm{R}}(k)\) within the USR inflation framework, we discern the amplification rate of \(P_{\mathrm{R}}(k)\) at its peak to be \(e^{6\Delta N}\).

The e-folding number of the end of the USR regime can be roughly settled in the range about $N\sim 15-50$ since inflation lasts for $60-70$ e-foldings. The USR regime may affect the CMB observable if $N$ is too small and the Universe fails to exit inflation if $N$ is too large. Given that $P_{R}\approx 2\times 10^{-9}$ at the CMB scales, the range of $\Delta N$ should be roughly $0-3.3$ to avoid nonlinearities of curvature perturbations. Our fitting results from PTA observations can also be understood as the constraints on $\Delta N$ in the range $N\sim 17.5-22.5$, shown in Fig.~\ref{fig:N-DeltaN posterior}. 
Although the current constraint is relatively loose, the future programs of GW observation, such as SKA, LISA, DECIGO and CE, can provide a much wider and more strict constraint on $P_{R}(k)$.

The comprehensive consideration of the CMB observable and the PTA results may yield some interesting results. Before the beginning of the USR regime, the rolling of inflaton tends to accelerate due to the steeper potential, rendering a more natural non-attractor phase. The early increase of $\dot{\phi}$ leads to a decrease of $P_{R}$ in the large scales. If the PTA data originates from the USR inflation, we expect a significant running of the spectral index of curvature perturbations in the more precise measurement of CMB. This explanation can also support double inflation~\cite{Silk:1986vc} where the inflaton exit the slow-roll regime during inflation and the effective potential has more than one plateau to provide enough e-folds.

To investigate the impact of \(\beta\) on constraints, we chose three typical values \(0.5,1,2\) for \(\beta\). The posterior distributions indicate that \(\Delta N\) strongly degenerates with \(N\). As mentioned before, \(\beta\) has little effect on the constraints.

	\begin{figure}[h]
		\includegraphics[width=0.5\textwidth]{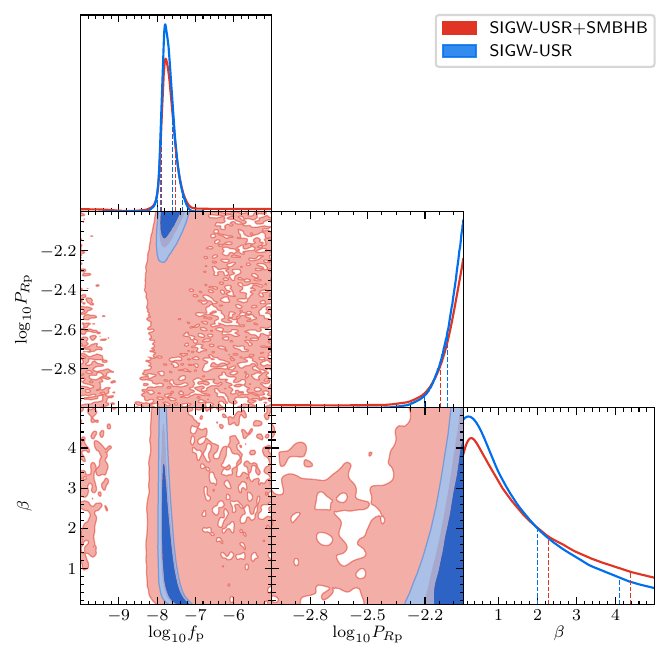}
        \includegraphics[width=0.5\textwidth]{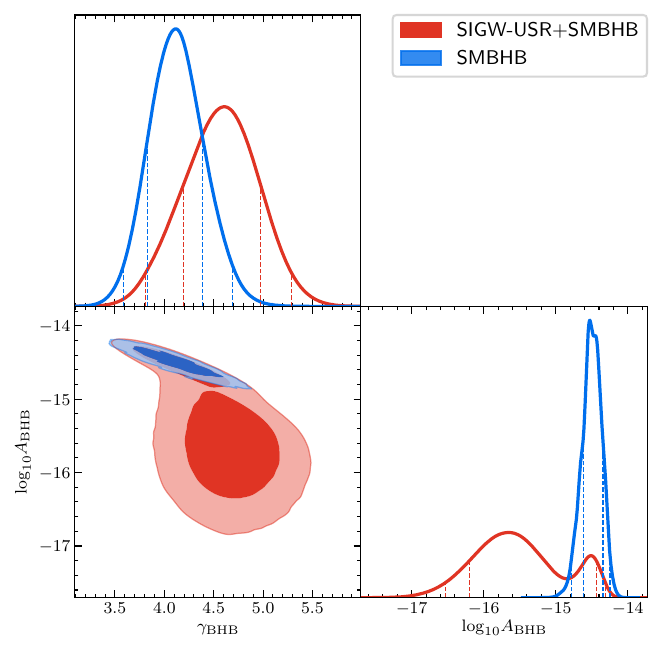}
		\caption{ Posterior distribution of three input parameters of SIGW-USR and SIGW-USR+SMBHB with the astrophysically motivated upper limit $\log_{10} P_{\mathrm{Rp}}\leq -2$, and of SMBHB model with the 2-D Gaussian prior.}
   
		\label{fig:posterior of usr and smbhb with astro priors}
	\end{figure}

\begin{figure}[h]
		\includegraphics[width=0.5\textwidth]{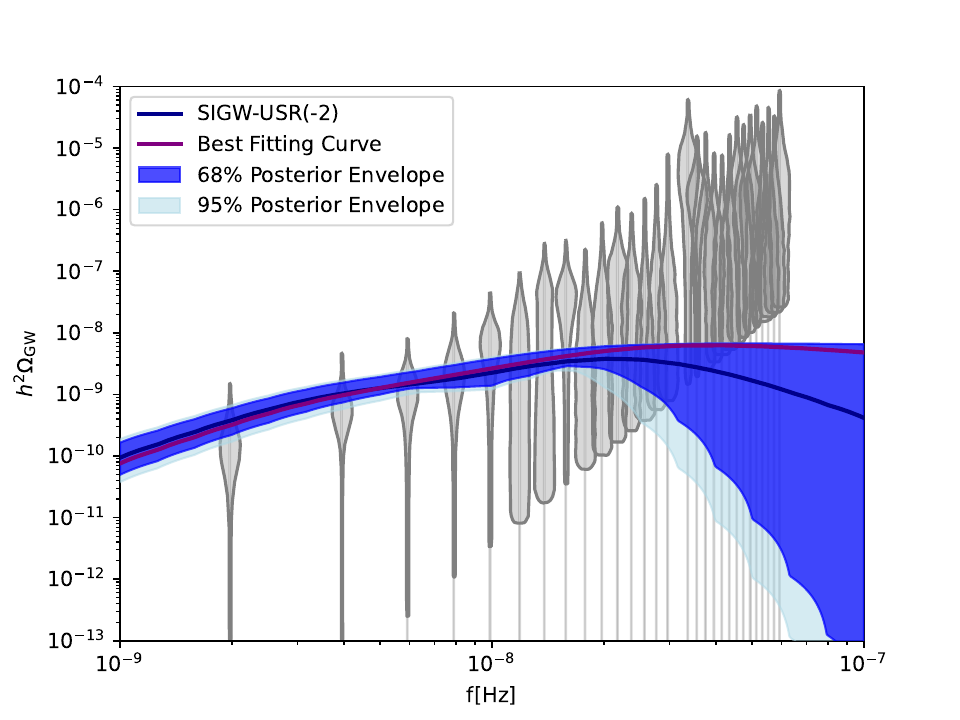}
        \includegraphics[width=0.5\textwidth]{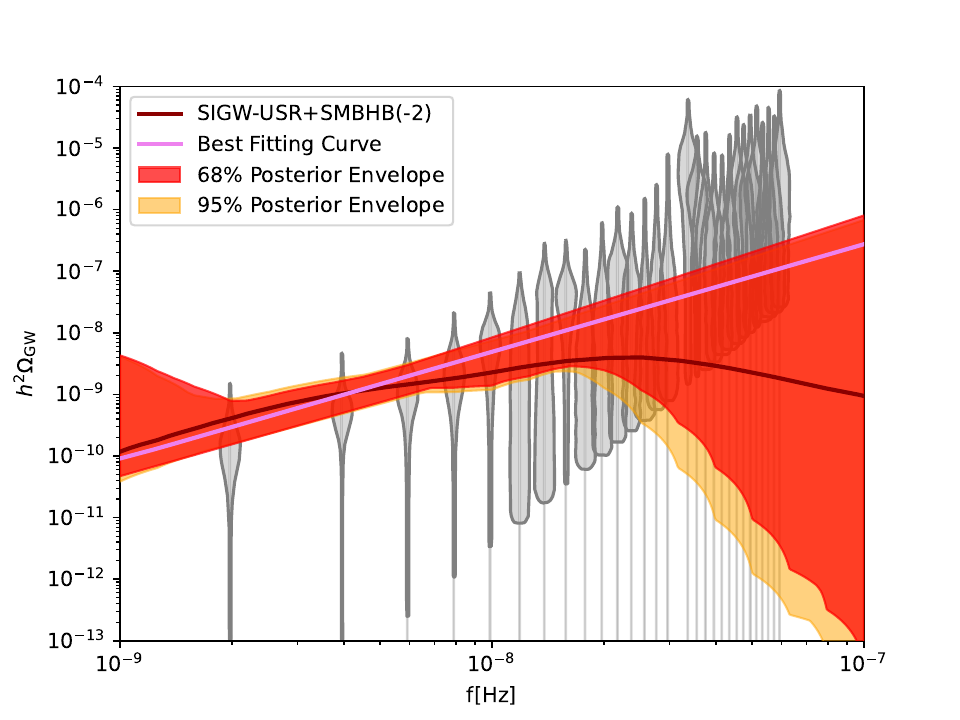}
        
        \includegraphics[width=0.5\textwidth]{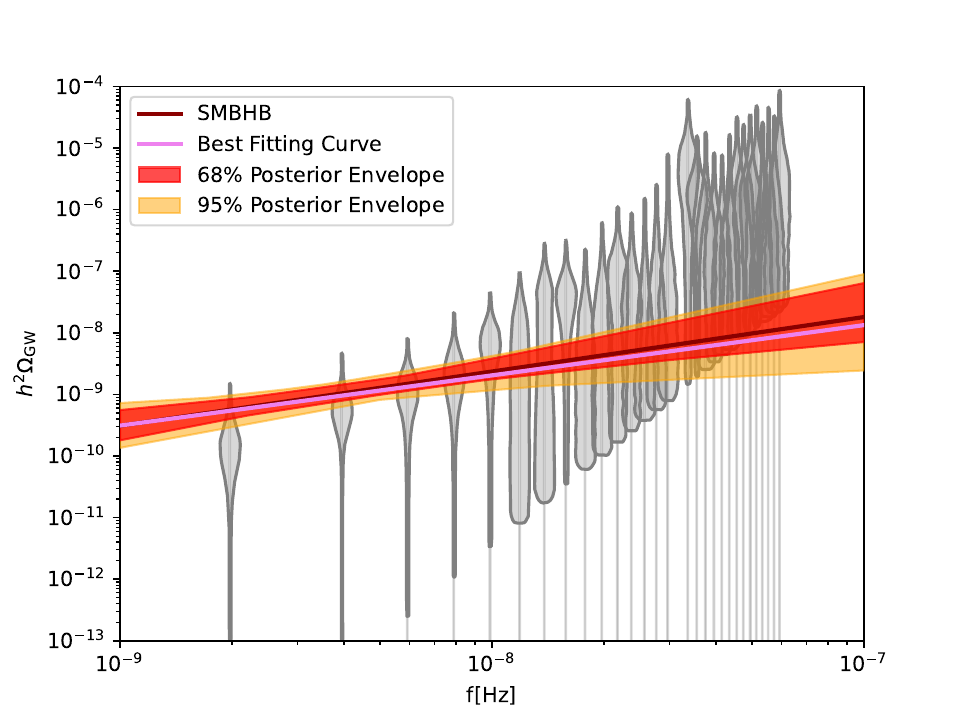}
		\caption{ The $68\%$ and $95\%$ posterior envelopes, the best-fitting SGWB power spectra, and the median SGWB power spectra for the SIGW-USR, SIGW-USR+SMBHB, and SMBHB models, all using astrophysical priors, are shown. The median GW power spectra labels are used to indicate which model each figure corresponds to. Since these figures don't show the $P_{\mathrm{Rp}}$ parameter directly, we mark (-2) to remind readers of the upper limit $\log_{10} P_{\mathrm{Rp}}\leq -2$.} 
		\label{fig:postenv of usr and smbhb with astro priors}
	\end{figure}

 To make a detailed comparison with the SMBHB model, Fig.~\ref{fig:posterior of usr and smbhb with astro priors} and Fig.~\ref{fig:postenv of usr and smbhb with astro priors} show the posterior distributions and posterior envelopes for the SIGW-USR, SIGW-USR+SMBHB, and SMBHB models, each with their astrophysically motivated priors. For the SIGW-USR model, the prior is $\log_{10} P_{\mathrm{Rp}} \leq -2$, while the SMBHB model uses the 2-D Gaussian prior described in Eq(\ref{2-D Gaussian parameters}). The Bayes factors against the SMBHB model are $\mathcal{B}=3.6\pm0.4$ for SIGW-USR and $\mathcal{B}=6.1\pm0.6$ for SIGW-USR+SMBHB, suggesting that, with astrophysically motivated priors, the NG15 dataset fits both the SIGW-USR and SMBHB models similarly well. This is further confirmed by Fig.~\ref{fig:postenv of usr and smbhb with astro priors}.

 In Fig.~\ref{fig:posterior of usr and smbhb with astro priors}, for the SMBHB model, the 2-D Gaussian prior introduces a smooth cutoff at the upper left end of the confidence contours, leaving some overlap with the $68\%$ confidence contour of Fig. 5 in Ref.~\cite{2306.16213}. In contrast, for the SIGW-USR model, the upper limit imposes a hard cutoff, thus the blue contours are extensions of the lower left end of the $95\%$ confidence contours in Fig.~\ref{fig:posterior of usr modeled GW}. The confidence contours for the SMBHB parameters in the SIGW-USR+SMBHB model represent a combination of the posterior distributions of the single SMBHB model and the 2-D Gaussian prior. The reasoning for this is introduced in the next paragraph.

 In Fig.~\ref{fig:postenv of usr and smbhb with astro priors}, the $68\%$ and $95\%$ posterior envelopes of the SIGW-USR+SMBHB model can almost be seen as a simple sum of those from the SIGW-USR and SMBHB models. This scenario occurs only when both models fit the NG15 dataset equally well. When models produce curves within their posterior envelopes, they are offering their best likelihoods under the given prior. In the case of SIGW-USR+SMBHB, if one model provides a curve within its own posterior envelope and the other model doesn't significantly detract from the performance, the best likelihood can still be achieved. As seen in the red contours of Fig.~\ref{fig:posterior of usr and smbhb with astro priors}, in the SMBHB posterior distributions, the upper $68\%$ contour corresponds to the SMBHB model dominating, while the lower $68\%$ contour corresponds to the SIGW-USR model dominating. The red $68\%$ contours also show some expansion, indicating that the combination of these two models enhances the performance of certain insufficient fits.

	\section{Conclusions and discussions}
   In this work, we explore the possibility of using USR inflation to explain the NG15 dataset and obtain the favored physical parameter space of the USR regime by the data. The posterior distributions indicate strong degeneracy between $\log_{10}P_{\mathrm{Rp}}$ and $\log_{10}f_{\mathrm{p}}$, and between $N$ and $\Delta N$. We obtain the $95\%$ C.L. lower limit $\Delta N > 2.8$ for $N\approx 20$, regardless of the value of $\beta$. In our previous work~\cite{Mu_2023}, the upper limit of $\Delta N$ given by the LIGO-Virgo O3 dataset is about $2.87$ for $N\approx 41.5$ with $\beta=1$.
   Compared to the LIGO-Virgo data, the NG15 dataset places strong constraints on $\Delta N$.

   In Fig.~\ref{fig:USR modeled GW with NANOGrav data}, we find that the best fitting spectra and the posterior envelopes from SIGW-USR alone and SIGW-USR+SMBHB are similar to each other. To compare the goodness of fit to the data for different models, one can calculate the Bayes factor. Our results reveal that the Bayes factors for the SIGW-USR model and the combined SIGW-USR+SMBHB model are comparable to the best-fit model reported in Ref.~\cite{NANOGrav:2023hvm}. Incorporating SMBHB leads to an increase in the Bayes factor. This enhancement can be attributed to the ability of the SMBHB spectrum to compensate for the insufficient fit of the SIGW-USR model alone, thereby enhancing the likelihood.

   For the power spectrum of curvature perturbations, we place the limit \( \log_{10} P_{\mathrm{Rp}} > -1.95\) at \(95\%\) C.L..
   Moreover, the one-dimensional posteriors of $\log_{10}P_{\mathrm{Rp}}$ and $ \log_{10}f_{\mathrm{p}}$ show the presence of peaks in the NANOGrav sensitivity band. The peaks can also be found in the SIGW-Gauss model in~\cite{NANOGrav:2023hvm} when $\Delta$ is small, where $\Delta$ is the scale parameter of the Gauss distribution characterizing the shape of $P_\mathrm{R}$.

   In Ref.~\cite{NANOGrav:2023hvm}, three SIGW models are investigated, mathematically modeled by different power spectrum $P_{\mathrm{R}}\left(k\right)$. The contours of $\log_{10} A - \log_{10}f_{\star}$ in ~\cite{NANOGrav:2023hvm} are quite similar with $\log_{10}P_{\mathrm{Rp}} - \log_{10}f_{\mathrm{p}}$ in this work. The reason for this is that the parameters in our model are similar to that of the SIGW models in~\cite{NANOGrav:2023hvm}, such as the infrared index, the peak amplitude, and the peak frequency. However, note that the energy spectrum in this work is 
   directly calculated from USR inflation. 

   To avoid the overproduction of PBHs, we apply the conservative upper bound, $\mathrm{log}_{10}P_{R\mathrm{p}}\leq -2$, for Gaussian curvature perturbations.~\footnote{$\Omega_{\mathrm{GW}}$ is also constrained by the observations of the effective freedom of relativistic particles from CMB experiments. The constraint turns out to be roughly $\Omega_{\mathrm{GW}}\lesssim 3\times 10^{-7}$ for the future CMB-S4 mission~\cite{CMB-S4:2016ple,Baumann:2015rya,Cang:2022jyc}, which already exceeds the GW strength observed by PTA experiments for about one order of magnitude. Therefore, considering the constraint from $\Delta_{\mathrm{eff}}$ has negligible impact on the results.} In this case, the SMBHB and SIGW-USR models fit the NG15 dataset almost equally well. In Fig.~\ref{fig:postenv of usr and smbhb with astro priors}, due to the 2-D Gaussian prior, the power-law cannot exhibit a steeper slope or larger amplitude, leading to poor performance in the lower frequency bins. For the SIGW-USR model, the upper limit on $P_{\mathrm{Rp}}$ forces the peak to occur early, resulting in a poor fit for the frequency bins to the right of $10^{-8}$ Hz. Although the combination of the best SGWB power spectra from these two models does not provide a significantly better fit to the NG15 dataset, this combination helps improve the poor performance of certain curves. However, as pointed out by many papers~\cite{Franciolini:2023pbf,Wang:2023ost,Liu:2023ymk,Yuan:2023ofl,Choudhury:2023fwk,Pi:2024jwt}, the primordial non-Gaussianities, even with a small value $f_{\mathrm{nl}}\lesssim -0.5$, can largely alleviate this problem and reduce the PBH abundance results from large curvature perturbations. 
   Primordial non-Gaussianities have been widely investigated as the method to detect the interactions of inflaton and distinguish other mechanisms that generating primordial curvature perturbations from CMB and large-scale structure observations~\cite{Bartolo:2004if,Chen:2010xka,Planck:2019kim}. The matter collapse into PBHs in the Hubble horizons where curvature perturbations exceed the threshold $\mathcal{R}_{c}$. Since $\mathcal{R}_{c}$ is generally more than $5$ times the standard deviation of curvature perturbations, a small correction on $\mathcal{R}_{c}$ can significantly enhance or depress the initial PBH abundance. The presence of non-Gaussianity can shift the threshold of PBH formation. The threshold can be lifted to suppress the PBH abundance with the help of a negative $f_{\mathrm{nl}}$, allowing a released upper bound on $P_{\mathrm{Rp}}$. Even a small value of $|f_{\mathrm{nl}}|$, is enough to make the USR model much more prefered by data. 
 The Bayes factor with the upper bound $P_{\mathrm{Rp}}<10^{-1.5}$ is already close to the case without upper any bounds on $P_{R\mathrm{p}}$.
   Therefore, we also obtain the result with $\mathrm{log}_{10}P_{R\mathrm{p}}\leq -1.5$, which indicates that the upper bound $\mathrm{log}_{10}P_{R\mathrm{p}}\leq -1.5$ captures most of the best-fit regions. In this case, the USR model is significantly preferred by data compared to the SMBHB model.

\begin{figure}[h]
        \centering
		\includegraphics[width=0.45\textwidth]{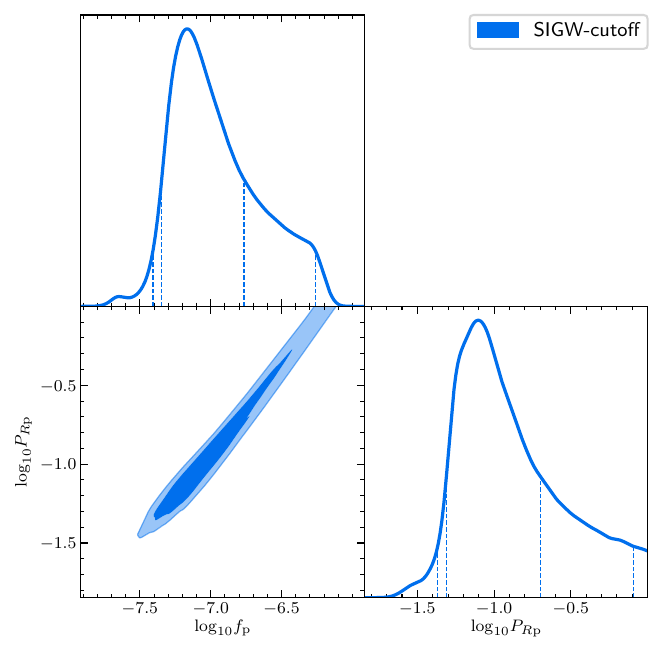}
		\caption{Posterior distribution of the parameters in the cutoff induced SIGW.
 The red and blue lines denote the $68\%$ and $95\%$ confidence regions, respectively.  }
		\label{fig:yd}
	\end{figure}

 We also provide the analysis of SIGWs from another type of $P_{\mathrm{R}}\left(k\right)$ which is induced by various GW sources after inflation~\cite{Liu:2022lvz,Zeng:2023jut,Papanikolaou:2020qtd,Domenech:2020ssp} and has a characteristic $k^3$ slope at the infrared side because of causality.
 At the ultraviolet side, we set a cutoff at the scale where the central-limit theorem becomes invalid.
 The posterior distribution is shown in Fig.~\ref{fig:yd}, which is quite similar to Fig.~\ref{fig:posterior of usr modeled GW}, since the precision of current data is not high enough to distinguish the slope $k^{3}$ and $k^{4}$. The result with a slope of $k^{3}$ agrees well with the previous one with a large $\beta$ since this cutoff model can be treated as an extreme USR model.

	\section*{Acknowledgments}
	We sincerely thank Zu-Cheng Chen for the helpful discussion. This work is supported in part by the National Key Research and Development Program of China Grant No. 2020YFC2201501 and No. 2020YFC2201502, in part by the National Natural Science Foundation of China under Grant No. 12075297, No. 12235019 and No. 12105060.

 \clearpage

 \begin{appendix}

     \section{The choice of $\gamma$ value}

In the parameterization of the power spectrum of curvature perturbations in USR inflation, the parameter $\gamma$ characterizes the smoothness of \(P_{\mathrm{R}}(k)\) around the peak. We compare our parameterization curves with the predictions of $P_{R}(k)$ from numerical simulations of many USR models and find that $\gamma=2.6$ is a representative parameter value.

\begin{figure}[h]
		\includegraphics[width=0.45\textwidth]{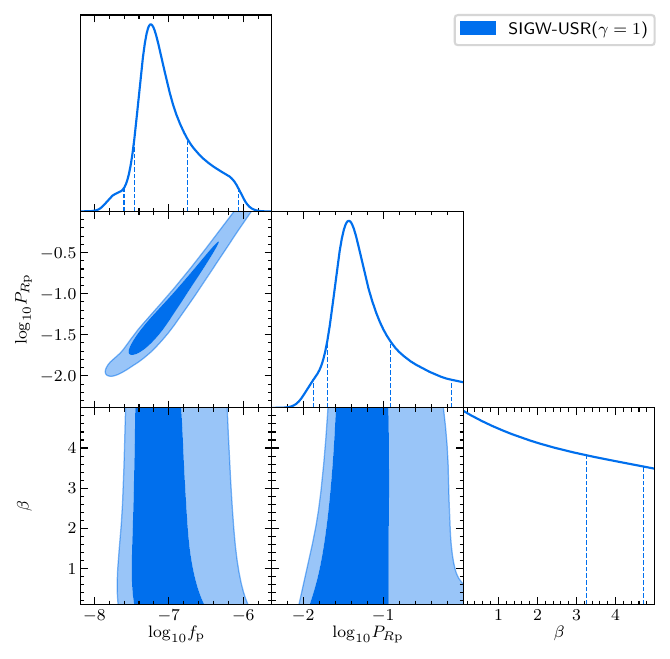}
        \includegraphics[width=0.45\textwidth]{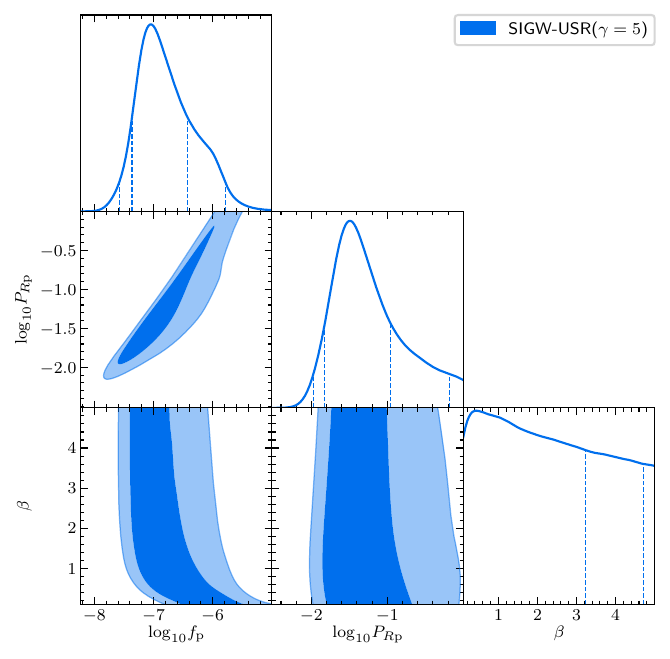}
		\caption{Posterior distribution of the parameters in the USR-SIGW model where $\gamma$ is chosen as $1$ (left panel) and $5$ (right panel).
 The red and blue lines denote the $68\%$ and $95\%$ confidence regions, respectively.  }
		\label{fig:c1_c5}
	\end{figure}

    Besides, we explore the influence of varying $\gamma$ values on the results. We employ \(\gamma=1\) and \(\gamma=5\) to fit the NG15 dataset, and the results are illustrated in Fig.~\ref{fig:c1_c5}. Comparing with Fig.~\ref{fig:posterior of usr modeled GW}, we find that the selection of \(\gamma\) has a negligible effect on the constraints. Only minor differences can be observed in the region of \(\beta<1\). With $\alpha$ fixed, large $\gamma$ and small $\beta$ lead to significant smoothing, causing a substantial increase in $P_{\mathrm{R}}(k)$. As a consequence, the peak frequency is expected to shift towards higher frequencies to match the observed data.

    Based on the above considerations, we fix \(\gamma=2.6\) in this paper.

 \end{appendix}

 \clearpage

 \bibliographystyle{apsrev}
	\bibliography{ms}

\end{document}